\documentstyle [12pt,preprint,eqsecnum,aps]{revtex}
\begin{document}

\draft

\title{Calculation of pure dephasing for excitons in quantum dots}
\author {Ehoud Pazy $^{1,2}$}
\address {$^1$ Ben - Gurion University of the Negev, Beer-Sheva 84105 Israel}
\address {$^2$ Institute for Scientific Interchange (ISI), Villa Gualino,
Viale Settimio Severo 65, I-10133 Torino, Italy}

\maketitle \begin{abstract}

Pure dephasing of an exciton in a small quantum dot by optical and acoustic
phonons is calculated using the ``independent boson model''. Considering the
case of zero temperature the dephasing
is shown to be only partial which manifests itself in the polarization decaying to a finite value.
Typical dephasing times can be assigned even though the spectra exhibits strongly non-Lorentzian 
line shapes. We show that the dephasing from LO phonon scattering, occurs on a much
larger time scale than that of dephasing due to acoustic phonons which 
for low temperatures are also a more efficient dephasing mechanism. The typical dephasing time is shown
to strongly depend on the quantum dot size whereas the electron phonon ``coupling
strength'' and external electric fields tend mostly to effect the residual
coherence. The relevance of the dephasing times for current quantum information processing implementation schemes in 
quantum dots is discussed.
 
\end{abstract}

\section {Introduction }
\label {sec:Introduction}
Semiconductor Quantum Dots (QDs) are of major interest for technological applications as well
as being a very interesting arena for basic physics. It has been predicted \cite{Sakaki82}
that through the use of QDs one could realize highly efficient lasers due to the QDs discrete 
density of states. Combining the QD technology and fabrication methods with ultra-fast laser
technology seems to be one of the most promising candidates for a realization of a quantum
information processing device (QIP). Ultra-fast laser technology allows one to produce and manipulate 
electron-hole quantum states on a sub-picosecond time scale, being the basis for potential 
implementations of quantum information/computation (QIC) \cite{Biolatti00,Reina00}, based on the excitonic degrees of freedom
in QD's. One of the most crucial requirements from a possible realization of a QIC
is the ability to perform coherent quantum manipulations. Thus a detailed computation and understanding 
of dephasing in QDs is badly needed. More-over it is highly important to obtain typical dephasing
times which can serve as limiting factors when considering coherent excitation of excitons
via laser pulses, i.e. the dephasing time is the quantity which defines 
the time scale on which coherent interaction of light and a quantum system can take place.

At low carrier densities in higher dimensional semiconductors structures such as 
bulk and quantum well the dephasing time is defined through
the scattering by LO phonons. Regarding QDs there is much controversy in the literature over 
phonon mediated relaxation process. The term ``phonon bottle neck'' \cite{Chow94} has been coined for 
the argument explaining why due to the discrete nature of the QD energy spectrum LO phonon can not lead to 
carrier relaxation if the LO phonon energy does not coincide with the energy level splitting. 
In the literature there is still much debate regarding the relevance of this argumentation \cite{Inoshita97}. 
In a series of papers by Verzelen and co-authors \cite{Bastard} it is claimed that electrons and LO phonons
are in the strong coupling regime leading to polaronic effects. When considering polaronic states
the arguments leading to ``phonon bottle neck'' effect are completely irrelevant, since LO
phonons mix electronic intra-bands in the QD, in their model relaxation is due to the finite life time 
of optical phonons. In this paper this controversial topic, i.e.,
the energy relaxation from excited exciton states to the excitonic ground state, shall not be
discussed, rather we are interested in the pure dephasing of the exciton ground state
at temperature equal to zero. The low temperature assumption will also allow us to ignore the polaronic
effects discussed above and we safely assume the electron and hole are in the QD ground state rather than some
polaronic superposition of intra-band states.
Pure dephasing \cite{Takagahara99} is the dephasing caused by virtual transitions
and is related to the decay of the off-diagonal terms in the exciton density matrix which does not lead
to a change in the occupation, i.e. the diagonal terms. Takagahara \cite{Takagahara99} treated pure dephasing 
due to acoustical phonons in a perturbative method including also excited exciton states. A recent very comprehensive study of the problem
of pure dephasing in small QDs was performed by Krummheuer and co-authors \cite{Khun02} which treated the problem exactly for
the excitonic ground state and have also considered the dephasing due to optical phonons. 
Whereas their treatment was numerical and focused on obtaining the absorption line shape in our work
we use the same approach but using more realistic wave functions, enables us to obtain some analytical results
and simple expressions for the typical dephasing time,
and remnant coherence.  

In Sec. \ref{sec:model} we will describe our model. In Sec. \ref{sec:optical} we calculate typical dephasing times
due to coupling to optical phonons and the remaining coherence in the Zero Phonon Line (ZPL), discussing two case:
strong electric field and no exterior field. It shall be shown that whereas the electric field does not strongly effect the
typical decoherence time, it effects the amount of remnant coherence. The decoherence effects of acoustic 
phonons will be reviewed in Sec. \ref{sec:accoustical} where it will be shown that the typical decohering 
time due to acoustical phonon is much shorter than that for optical phonons, and that for low temperatures
acoustic phonons more efficient in decohering excitons. For a weak external electric field the deformation
potential mechanism will be shown to be more efficient in dephasing than the piezoelectric but for a strong
external electric field they will be shown to be comparable. A short summary will be given in Sec. \ref{sec:Summary}
and the relevance of this work to the implementation of QIP will be described.

\section{The model}
\label{sec:model}
We consider a GaAs-based structure where the confining potential for the QD's is modeled by
a 2D harmonic potential, defined by the frequencies $\omega_{e,h}$ for the electron and hole in the $(x,y)$-plane
and square-like along $z$-the growth direction. This simple model for the confining potential 
seems to capture much of the important physics of In$_{x}$Ga$_{1-x}$  QDs as well as proving
to be reliable in analyzing experimental data. It should be noted that the model we use is more realistic than the
one used by Krummheuer and co-authors \cite{Khun02} which considered an infinite well confining potential
in the growth direction. The QDs we consider are self-assembled InAsGaAs
QDs (as in the model of QIC considered in Ref. \cite{Biolatti00})
thus typical parameters, e.g., electron hole effective masses, phonon sound velocity, will be taken 
as GaAs bulk values. We will also consider the same model with an
electric field, $F$, in the $x$-direction. Such an electric field can be external or due to a 
constant strain field such as the one appearing in GaN QDs \cite{re:gan}. The Hamiltonian containing the 
electric field term is given by:
\begin{equation}
H^{e,h}({\bf r})=- \frac{\hbar^2 \nabla_{\bf r}^2}{2 m_{e,h}} +V^{e,h}_0 \theta(|z|-{L \over 2})+
{m_{e,h} \omega_{e,h}^2 \over 2} (x^2+y^2) \mp e F x,
\end{equation}
where the $-/+$ signs refer to the electron and hole respectively, 
subscripts and superscripts ${e,h}$ will be used hereafter to denote electron/hole related quantities,
$m_{e,h}$ is the effective mass of the electron (hole), $V_{0}$ is the square potentials 
height, $L$ is it's width, $\theta$ is the step function and $e$ is the electric charge.
Taking the electric field in the $x$-direction just causes a relative shift by 
$x_{0}= e F ({1 \over m_{e,h} {\omega_e}^2}+{1 \over m_{e,h} {\omega_h}^2}) $ of the confining harmonic
potential (as well as a constant energy shift).

We consider small QDs, i.e. the strong confinement regime, which leads to well separated
sub-levels within the QD. This allows us to safely treat only the electron-hole pair ground state
level and to neglect Coulomb interaction between the electron and hole, which is small in
${L \over R_{B}}$, where $R_{B}$ is the bulk exciton Bohr radius.

In the strong confinement regime the Hamiltonian for the exciton coupled to phonons can be written as \cite{Rink87}
\begin{equation}
\label{eq:hamiltonain}
H= \epsilon_{ex}c^\dagger c + \sum_{n{\bf q}} \hbar \omega_{n {\bf q}}
(b_{n{\bf q}}^\dagger b_{n{\bf q}} +{1 \over 2}) + 
\sum_{n {\bf q}}c^\dagger c (M_{n{\bf q}} b_{n{\bf q}} + H.c.), 
\end{equation} 
where $\epsilon_{ex}$ is the exciton energy which is a sum of the electron and hole 
ground state energies, $c^\dagger$ ($c$) ,are exciton, creation 
(annihilation) operators respectively and $b_{n{\bf q}}^\dagger$($b_{n{\bf q}}$) are creation 
(annihilation) operators for phonons in branch $n$ and wave number $q$ with corresponding
energy of $\hbar \omega_{n {\bf q}}$.
$ M_{n{\bf q}}$ the exciton phonon matrix element is given by
\begin{equation}
\label{eq:matrix}
M_{n{\bf q}}=M_{n{\bf q}}^c \int d^3 r \ e^{\imath {\bf q r}} | \phi^e ({\bf r}) |^2 -
M_{n{\bf q}}^v \int d^3 r \ e^{\imath {\bf q r}} | \phi^h ({\bf r}) |^2 .
\end{equation}
Here $\phi^{e,h}$ denotes the electron and hole
envelope functions respectively and $M_{n{\bf q}}^c$ ($M_{n{\bf q}}^v$) is the conduction
(valance) electron (hole)-phonon matrix element in the bulk. In order to simplify notations
we will hereafter suppress the phonon branch index $n$.

\subsection {Optical susceptibility}
\label {subsec:susceptability}

The Hamiltonian (\ref{eq:hamiltonain}) (``independent boson model'') can be diagonalized exactly 
\cite{Mahan90}. Being interested in the low temperature regime we will consider the zero 
temperature limit, under this assumption the imaginary part of the optical susceptibility is given by
\begin {equation}
\label{eq:susceptability}
Im \chi (\omega) = {C\over 2 \pi  V \hbar} \int_{-\infty}^{\infty} dt 
\exp{ \left [ {\imath t \over \hbar}
\left (\omega - \epsilon_{ex} \right) - 
\sum_{\bf q} {|M_{\bf q}|^2 \over \hbar^2\omega_{\bf q}^2} 
\left [ 1 - e^{- \imath \omega_{\bf q} t}\right ] +
\imath  \sum_{\bf q}{|M_{\bf q}|^2 \over \hbar^2 \omega_{\bf q}} t 
 \right]}
\end{equation}
where $V$ is the normalization volume and $C={2 \pi e^2 \hbar^2 |p_{cv}|^2 \over m_{0}^2 \omega^2}$ is considered as a constant
\cite{Rink87}, i.e. neglecting variations in the photon energy $\omega$ compared to $ \epsilon_{ex}$. In estimating the
prefactor, $C$, the difference between the electron and hole effective masses was also neglected. 
Denoting the bulk optical matrix element by $p_{cv}$, the free electron mass by $m_{0}$.

The relationship between the imaginary part of the susceptibility and the dephasing which is defined
through the polarization relaxation time is evident when one considers the linear response regime.
In the linear response regime the polarization is proportional to the applied electric field, $E$ in 
following way, $P(\omega)=\chi(\omega)E(\omega)$. If one considers the electric field to be switched
on as a $\delta$ shaped pulse the imaginary part of the polarization is proportional to the dissipation.
The remaining coherence which we refer to, manifests itself through the decay of the polarization to
a finite non-zero value.
It should be mentioned that in restricting ourselves to the linear response regime we consider weak laser pulses, i.e.
small Rabi frequencies. Typically one can expect that quantum information processing will be implemented
in the strong field regime, but all one has to do in that case is to rewrite the current discussion
in terms of laser dressed states \cite{Calarco}.

In order to estimate the dephasing time one has to calculate the exciton-phonon matrix elements.
In order to calculate these matrix elements, $M_{n{\bf q}}$, we use the following 
simplifying assumption: we approximate the electron's and hole's wave function in the $z$ (growth)
direction by a Gaussian wave function (see Ref.\cite{Damico01} ). In this approximation the form factors 
are given by
\begin{equation}
\label{eq:fourier}
{\cal F}_{\bf q}^i=\int d^3 r \ e^{\imath {\bf q r}} | \phi^i ({\bf r}) |^2 =
\exp{\left [
-{1 \over 2}( q^2 \sigma^{(i)} + q_{z}^2 \sigma^{2(i)}_{z}) 
\right ]},
\end{equation}
where $q = \sqrt{q_{x}^2 + q_{y}^2}$ and 
$ \sigma^{(i)}$,$\sigma_{z}^{(i)}$  give the spatial scale of the wave function in the plane and
the growth direction respectively using the superscript
$i={e,h}$ to denote the electron and hole wave functions respectively.
The spatial scale for the wave function in the in-plane direction is given by the width of the Gaussian
(the confining dot potential is parabolic) wave function, $\sigma^{(i)}=\sqrt{\hbar \over m^i \omega^i}$.
In the growth direction we approximate the spatial scale of the wave function by 
$\sigma_{z}^{(i)}=\sqrt{\langle z^2 \rangle}$, the expectation value calculated by taking the ground
state wave function of an infinite potential well with width $L$.
We simplify the calculations further by considering an isotropic electron(hole)-phonon coupling
for all coupling mechanisms, i.e. Fr\"{o}lich coupling to optical phonons
 and piezoelectric and deformation potential coupling to acoustic phonons.
\section{Optical phonons}
\label{sec:optical}
Typically the polar coupling of the exciton to optical phonons tends to
be neglected \cite{Rink87} since for infinite confining barriers the matrix element (\ref{eq:matrix}) is zero
(if one assumes the same effective mass for the electron and hole) due the cancellation of the 
electron and hole ``polaron clouds''. In considering finite confining
potentials for electrons and holes, i.e. using different wave functions for electrons and holes due to 
their different effective masses and confining potentials, one can obtain enhanced exciton polar phonon interaction\cite{Heitz99}. 

The exponential term in Eq. (\ref{eq:susceptability}) which is responsible for the dephasing is 
$ \Phi(t) \equiv \sum_{\bf q} {|M_{\bf q}|^2 \over \hbar^2\omega_{\bf q}^2} 
\left [ 1 - e^{- \imath \omega_{\bf q} t}\right ]$ and the last
term in that equation: $\imath  \sum_{\bf q}{|M_{\bf q}|^2 \over \hbar^2 \omega_{\bf q}} t$,
is just the polaronic shift which in the case of optical phonons is just a constant renormalization of the 
exciton's energy due to the interaction with phonons. 
Using the standard Fr\"{o}lich coupling for electron(hole)-LO-phonon interaction
$M_{\bf q}^{op} = {A \over V^{1 \over 2} q}$ (the superscript $op$ denotes the optical-phonon branch), where 
$A^2 = {\hbar \omega_{q} e^2 \over 2 \epsilon_{0}}   
\left ( {1 \over \kappa_{\infty}}- {1 \over \kappa_{0}} \right )$, 
$ q= |{\bf q}|$, $ \kappa_{\infty}$ and $ \kappa_{0}$ are high frequency and static dielectric 
constants $\epsilon_{0}$ is the vacuum susceptibility. Replacing the summation over ${\bf q}$ by integration 
${1\over V} \sum_{\bf q} (...)= \int {d^3q \over (2 \pi)^3} (...)$ the expression for $\Phi^{op}(t)$ is
\begin{equation}
\label{eq:dephop}
\Phi^{op}(t)= {B \over {(2 \pi)}^3 \hbar}\int_{-\infty}^{\infty} d q_{z} \int_{0}^{\infty} dq  K(q) {1 -e^{- \imath \omega_{q} t} 
\over \omega_{q}} ,
\end{equation}
where $B={e^2 \over 2 \epsilon_{0}}  \left ( {1 \over \kappa_{\infty}}- {1 \over \kappa_{0}} \right )$ \\
and $K(q)= {2 \pi q \over  {q^2 + q_{z}^2}} 
\left ( e^{- (q^2 {\sigma^e}^2+ q_{z}^2 {\sigma_{z}^e}^2)} + 
e^{- (q^2 {\sigma^h}^2 + q_{z}^2 {\sigma_{z}^h}^2)} - 
2J_{0}(q x_{0})e^{- [{q^2 \over 2} ({\sigma^e}^2 + {\sigma^h}^2) +
{q_{z}^2\over 2}( {\sigma_{z}^e}^2 + {\sigma_{z}^h}^2)]} \right )$, \\ 
and $J_{0}$ is the zeroth order Bessel function. The three terms in $K(q)$
corresponding to the electron, hole and the overlap of the electron and hole, contributions. 
The stronger the field the less contribution we get from the overlap term, in the opposite 
limit: no field, the over lap term exactly cancels the electron and hole contributions, if one 
assumes the same wave function for the electron and the hole.

Without introducing dispersion for the optical phonons, i.e. for Einstein phonons with frequency
$ \omega_{0}$, we get
$\Phi(t)=S \left(1 - e^{- \imath \omega_{0} t} \right)$ where 
$S ={B \over {(2 \pi)}^3 \hbar}\int_{-\infty}^{\infty} d q_{z} \int_{0}^{\infty} d q 
{K(q) \over \omega_{0}} $. This leads to phonon side-bands or satellites, 
the spectral function is composed of a sum of delta functions evenly spaced with a separation
of $\omega_{0}$ with a decreasing amplitude \cite{Rink87,Mahan90}.

Introducing optical phonon dispersion relations
(for details see Appendix \ref{ap:dispersion}) one can expand $\omega_{q}$ around $\omega_{0}^{op}=\sqrt{2 c_{1}^{op}}$
(which is the value of $\omega_{q}$ for $q=0$)
getting the following expression for $\Phi(t)$ 
\begin{equation}
\label{eq:disexp}
\Phi^{op}(t)=
{B \over {(2 \pi)}^3 \hbar}\int_{-\infty}^{\infty} d q_{z} \int_{0}^{\infty}
{K(q) \over \omega_{0}^{op}} 
\left [1 - \exp \left ( -\imath \omega_{0}^{op}t + \imath f \ q^2 t\right ) \right ] \ ,
\end{equation}
where $ f \equiv {c_{2}^{op} \over c_{1}^{op} {\omega_{0}^{op}}}{\left ({ \pi  \over 4 k_{max}} \right )}^2 $ which is proportional
to the second derivative of $\omega(q=0)$ and acts 
as a measure of the deviation of the optical phonon band from flatness. Parameters: $c_{1}^{op}, c_{2}^{op},k_{max}$
are defined in Appendix \ref{subsec:optical}.

Integrating Eq. (\ref{eq:disexp}) we obtain (for details see Appendix \ref{subsec:optical}) for the 
case of no electric field, i.e. $x_{0}=0$
\begin{eqnarray}
\label{eq:nofield}
\Phi^{op}(t) & = &
\sum_{j={e,h}} P_{j}
\left [
\ln {\left ( {p_{j} +v_{j} \over p_{j} - v_{j}} \right )}
- \ln {\left ( {\sqrt{p_{j}^2 - {\imath f t \over{\sigma^j}^2} } + v_{j} \over 
\sqrt{p_{j}^2 - {\imath f t \over{\sigma^j}^2} } - v_{j}} \right )} 
e^{-\imath \omega_{0}^{0p} t}
\right ] \nonumber \\
& -& \bar{P}\left [
\ln {\left ( {\bar{p} + \bar{v} \over \bar{p} -\bar{v}} \right )}
- \ln {\left ( {\sqrt{\bar{p}^2 - {\imath f t \over \bar{\sigma}^2} } + \bar{v} \over 
\sqrt{\bar{p}^2 - {\imath f t \over \bar{\sigma}^2} } - \bar{v}} \right )} 
e^{-\imath \omega_{0}^{0p} t}\right ],
\end{eqnarray}
where $P_{j}={B \over {8 \pi^2} \hbar}
{1 \over \omega_{0}^{op}}{\sqrt{\pi} \over \sigma_{j} v_{j}} $,
$\bar{P}={B \over {4 \pi^2} \hbar}
{1 \over \omega_{0}^{op}}  
{\sqrt{\pi} \over \bar{\sigma}\bar{v}} $, a measure of the spatial extent of the wave functions is given by
$\bar{\sigma}^2 \equiv \ {\sigma^e}^2 +{\sigma^h}^2 $,
 $\bar{\sigma}_{z}^2 \equiv {\sigma_{z}^e}^2 +{\sigma_{z}^h}^2$ and the anisotropy of the wave functions is given via
$p_{j}={\sigma_{z}^{j} \over \sigma^{j} }$, $\bar{p}={\bar{\sigma}_{z} \over \bar{\sigma}}$, 
$v_{j}=\sqrt{p_{j}^2- 1}$, $\bar{v}=\sqrt{\bar{p}^2 - 1}$.

It is easily checked that for the case of identical wave functions for electron and hole, i.e. 
$\sigma^e = \sigma^h$ and $\sigma_{z}^e = \sigma_{z}^h$ one obtains that $\Phi(t)$ is identically
zero. \\
For the case of a strong electric field \cite{re:opt}: when the separation between the electron and hole wave 
function being much bigger than the typical scale of the wave functions , i.e. 
$x_{0} \ll \sigma^j,\sigma_{z}^j$  $j={e,h}$, we obtain using a saddle point approximation
\begin{eqnarray}
\label{eq:strongfield}
\Phi^{op}(t)& = &
 \sum_{j={e,h}} P_{j} 
\left [
\ln {\left ( {p_{j} + v_{j} \over p_{j} - v_{j} } \right )}
- \ln {\left ( {\sqrt{p_{j}^2 - {\imath f t \over{\sigma^j}^2} } + v_{j} \over 
\sqrt{p_{j}^2 - {\imath f t \over{\sigma^j}^2} } - v_{j}} \right )} 
e^{-\imath \omega_{0}^{0p} t} \right ] \nonumber \\
& - & \bar{P} {\sqrt{2} \bar{\sigma}\bar{v} \over x_{0}^2} 
\left [ \bar{\sigma_{z}} - 
\sqrt{(\bar{\sigma_{z}}^2 +\imath f t)}
e^{-\imath \omega_{0}^{0p} t}\right ],
\end{eqnarray}

Typical decoherence times for the electron, hole can be directly obtained
by comparing the coefficients of $q^2$ in Eq. (\ref{eq:disexp}) 
or by considering the result of the integration Eq. (\ref{eq:nofield}) giving:
$\tau_{dph}^e \simeq {{\sigma^e}^2-{\sigma_{z}^e}^2 \over f}$, 
$\tau_{dph}^h \simeq {{\sigma^h}^2-{\sigma_{z}^h}^2  \over f}$.
Using the typical QD parameters considered in Ref. \cite {Biolatti00}:
in which the QD width is $L=50 \AA~$, typical energy scales for parabolic confinement are 
$\hbar \omega^e = 30 meV$ for the electron and $\hbar \omega^h = 24 meV$ and the parameters for phonon 
dispersion relations for GaAs can be found in Appendix \ref{ap:dispersion}, we obtain decoherence times of the order of:
$\tau_{dph}^e \simeq 200 ps$ and $\tau_{dph}^h \simeq 50 ps$. 

\subsection {Residual Coherence}
\label {subsec:coherence}
The residual polarization does not decay in our model,
rather one needs other decoherence mechanisms for it's decay such finite optical phonon
life time \cite{uskov00}. The residual coherence left in the ZPL is given by:
$exp{[-\Phi (t \longrightarrow \infty)]}$. It is again instructive to review the case of Einstein 
phonons \cite{Mahan90}. Calculating the spectral function in the Einstein model gives
$A(\omega)= 2 \pi e^{-S} \sum_{l=0}^{\infty} {S^l \over l!} 
\delta(\omega-\epsilon_{ex}+S\omega_{0}-\omega_{0}l)$, 
where $\omega_{0}$ is the phonon frequency.
The spectral function satisfies the following sum rule: 
$1=\int {d \omega \over 2 \pi} A(\omega)= e^{-S}\sum_{l=0}^{\infty} {S^l \over l!}$.
In our model this sum rule is also trivially satisfied for $t=0$ (since $\Phi (0)=0$), but as time
proceeds the terms containing $l\neq 0$ decay. The decoherence in our case 
should be considered more as a spectral diffusion phenomena related to the reduction 
of the exciton quasiparticle spectral weight which is effected without
any additional resonance broadening (for a similar effect obtained in QDs in the Kondo regime see Ref. \cite{Silva01}). 

The remaining coherence in our model for the case of coupling to optical phonons is given by $e^{-\tilde{S}}$ where $\tilde{S}$ is the
Huang-Rhys parameter. A simple expression can be obtained using the typical anisotropy of 
the confining potential in self-assembled QD's in which the confinement along the growth direction, $z$, 
is normally much stronger than the confinement along the in-plane direction thus generally 
$({\sigma_{z}^j \over \sigma^j})^2 \ll 1$ where $j={e,h}$. In the case of a strong electric field 
$30 \  kV/cm$ one can completely neglect the contribution from the overlap part in 
Eq. (\ref{eq:strongfield}) in the calculation 
of remaining coherence, since this term is smaller by a factor of  $({\sigma_{z} \over  x_{0}})^2\ll 1$, 
(in our case ${\sigma_{z} \over  x_{0}}\simeq 0.2$). 
Therefore we obtain
\begin{equation}
\tilde{S}= {B \over 4 \pi \sqrt{\pi} \hbar}{1 \over \omega_{0}^{op}} 
( {\varphi_{e} \over \sigma^{e}} +  {\varphi_{h} \over \sigma^{h}}),
\label{eq:strgfield}
\end{equation}
where $\varphi$ is defined through $\tan (\varphi_{j}) = 2p_{j}$, 
having neglected terms of order $p_{j}^2$. From the values of the parabolic confinement potentials for electrons and
holes, using the following effective masses $m_{e}=0.067 m_{0}$ and  $m_{h}=0.34 m_{0}$ for electrons and 
holes respectively, we get $\sigma^{e}=54\AA~$ and $ \sigma^{h}=27 \AA~$. Through the width of the quantum dot
one obtains $\sigma_{z} =17\AA~$.
Using these values and the dielectric constants of GaAs, $\kappa_{\infty}=10.7$ and $\kappa_{0}=12.4$,
we obtain for the case of a strong electric field, $\tilde{S}=0.033$. 

For the case of no external electric field all terms are of the same order and the overlap term nearly cancels 
the sum of the electron and hole parts, this expresses the fact that the electron and hole couple much weaker 
to phonons (due to local charge neutrality) resulting in a large remnant coherence,
\begin{equation}
\tilde{S}= {B \over  4 \pi \sqrt{\pi} \hbar}{1 \over \omega_{0}^{op}} 
( {\varphi_{e} \over \sigma^{e}} +  {\varphi_{h} \over \sigma^{h}} - 2 {\bar{\varphi} \over \bar{\sigma}} ),
\label{eq:strongfi}
\end{equation}
where $\tan (\bar{\varphi})=2\bar{p}$, having neglected terms of order $\bar{p}^2$.
In the case of no external electric field $\tilde{S}$ comes out to be $\tilde{S}=0.008$. 
It is interesting to compare this value to 
experimental and theoretical values of Huang-Rhys parameter quoted in the literature for pyramidal 
InAs/GaAs QDs \cite{Heitz99}, $S \approx 0.01$ (for a QD with a pyramid baseline of $\approx 170 \AA~$).
The last values are comparable to our results and the difference can be attributed to different 
electron and hole wave functions.

\section{Acoustical phonons}
\label{sec:accoustical}

Pure dephasing by acoustical phonons becomes more dominant with the rise of temperature. At low temperatures
the effect of pure dephasing due to scattering with acoustical phonon is small since there are not many
real phonons to scatter with and the main contribution comes from virtual processes. For optical phonons the
zero temperature limit is defined by the temperature  bellow one can safely ignore polaronic effects, i.e.,
the mixing of electronic intra-bands in the QD induced by optical phonons (see Sec. (\ref{sec:Introduction})). 
For acoustic phonons one needs to go to temperatures bellow $2 K$ (for details see 
Appendix \ref{subsec:apdef}).

We will consider two different exciton acoustic-phonon coupling mechanisms: deformation potential and
piezoelectric. The calculation of the dephasing is again reduced to calculating the exciton phonon matrix elements.

\subsection {Deformation potential}
\label {subsec:deformation}
In the case of zero electric field the deformation potential is expected to be the leading
dephasing mechanism since the deformation potential coupling constant is different for 
electrons $D^{c}=-14.6eV$ and holes $D_{v}=-4.8eV$. Therefore even in the case when the
electrons and holes have the same wave function still there will be dephasing due to
the deformation potential coupling, which is not the case for the piezoelectric coupling
mechanism.

The bulk matrix element for the deformation potential coupling is given by
$M_{a{\bf q}}^{c/v} = {\left (\hbar q \over 2 V \rho u \right )}^{1/2} D$, \ 
(the subscript $a$ denotes the acoustic-phonon branch and the superscript denotes the conduction
(valance)  band values) $\rho=5.3 g/cm^3$, is the crystal density 
and $u=4.8 \times 10^5 cm/s$ is the
phonon (longitudinal) sound velocity. Replacing the summation over ${\bf q}$ by integration 
the expression for $\Phi_{ac}^{def}(t)$ is
\begin{equation}
\label{eq:depexp}
\Phi_{ac}^{def}(t)= {C_{d} \over {(2 \pi)}^3 \hbar}\int_{-\infty}^{\infty} d q_{z} \int_{0}^{\infty}  dq
\sqrt{q^2 +q_{z}^2}\ K(q) 
 (1 -e^{- \imath \omega_{q} t}) ,
\end{equation}
where $C_{d}={D^2 \over 2 \rho u^3}$.

Performing the integration via a saddle point approximation (for details see Appendix \ref{subsec:apdef})
we obtain for $t \gg {p \over u}$, for the case with no external electric field  ($x_{0}=0$) 
\begin{eqnarray}
\label{eq:dephasing}
\Phi_{ac}^{def}(t) & = & 
\sum_{j={e,h}} A_{j}
\left\{
\tanh^{-1}{\left( s_{j} \right)} 
 - \pi e^{- {\left (t \over \tau_{j} \right ) }^2} 
Erf{ \left ({t \ s_{j} \over \tau_{j} }
\right )}
\right \} - \nonumber \\
& - & \bar{A}
\left\{
\tanh^{-1}{\left( \bar{s} \right)} 
 - \pi e^{- {\left (t \over \bar{\tau} \right ) }^2} 
Erf{ \left ({t \bar{s} \over \bar{\tau}} 
\right )}
\right \}
\ ,
\end{eqnarray}
where, $A_{j}={C_{d}^{j} \over {\sigma^{j}}^2 s_{j}(2 \pi)^2 \hbar}$,
$\bar{A}={2 \bar{C}_{d} \over {(2 \pi)}^2 \hbar}{1 \over \bar{\sigma}_{j}^2 \bar{s}}$,
$\tau_{j}= {2 \sigma_{j}\over u}$, $\bar{\tau}={2 \sigma_{j}\over u}$,
$s_{j}=\sqrt{1- p_{j}^2}$, $\bar{s}=\sqrt{1- \bar{p}^2}$,
$\bar{C}_{d}= {D_{e}D_{h} \over 2 \rho u^3}$
and  $Erf(z)= {2 \over \sqrt{\pi}} \int_{0}^{z} e^{-t^2}$, is the error function.

Estimating the decoherence time using $\tau_{dph}^j \simeq {2 \sigma^{j} \over u}$ we obtain for the 
electron and the hole typical decoherence times of the order of several pico-seconds. To estimate 
the remaining coherence we first consider the case of a strong electric field. As in 
the case of the optical phonons one could again calculate the overlap term using the saddle point
method. The overlap term would be much smaller then the electron and hole terms by a factor of
${\sigma \over x_{0}}^{2}$ to some power. Thus the remaining coherence 
will be given by the sum of electron and hole terms:
$e^{-g_{def}}$, where $g_{def}= \Phi^{def} (t \longrightarrow \infty)] \approx g_{def}^{e}+g_{def}^{h}$
where $g_{def}^{j}={C_{d}^{j} \over (2 \pi \sigma^{j})^2 \hbar}{\tanh^{-1}{\left( s^j
\right)} \over s^j} $. For the same parameters used for the calculation of the remaining coherence
for optical phonons (see \ref{subsec:coherence}) we obtain $s^{e}=0.95$ giving
$g_{def}^e = 0.032$ and for the hole $s^h=0.78 $ giving $g_{def}^h= 0.01275$ and $g_{def}=0.04475$. These values are larger but still
comparable to the values obtained by the coupling to optical phonons, for the case of a strong field.
When considering the case of no external electric field, $g_{def}$ is reduced to
$g_{def}=0.022$, due to the overlap term. As can be expected the deformation potential is much less
sensitive to a strong electric field compared to the sensitivity of optical polar coupling.

\subsection {Piezoelectric potential}
\label {subsec:piezoelectric}
In assuming that the electrons and holes have the same wave functions and they only
differ through there different effective masses, as was done in Ref. \cite{Khun02},
for small electric fields one can safely ignore the dephasing due to the piezoelectric coupling
mechanism due to the large overlap of hole and electron wave function, i.e., very small dipole
moment. In contrast since we assume different wave functions for the electron and hole so we obtain
quite substantial (comparable to the deformation potential) dephasing due to the piezoelectric
coupling even in the zero field case. For the strong electric field case the dephasing due to the
piezoelectric coupling mechanism is larger in our case than the one induced by the deformation potential
coupling. The bulk matrix element for the piezoelectric potential is given by
$M_{a{\bf q}}^{c/v} = {e e_{14} \over \epsilon_{0} \epsilon_{s}}
 \sqrt{\hbar \over 2 V \rho u |q|}$, where $e_{14}=0.16 C/m^2$ is the piezoelectric coupling constant  
(having assumed a spherically symmetric model). The expression for $\Phi_{ac}^{pz}(t)$ is
\begin{equation}
\label{eq:dephpiezo}
\Phi_{ac}^{pz}(t)= {C_{p} \over \hbar}\int_{-\infty}^{\infty} d q_{z} \int_{0}^{\infty}  dq
{K(q) \over \sqrt{q^2 +q_{z}^2}}  
 (1 -e^{- \imath \omega_{q} t}) ,
\end{equation}
where $C_{p}= {e_{14}^2 \over 4 \pi^2 \epsilon_{0}^2 \epsilon_{s}^2 } 
{e^2 \over 2 \rho u^3}$ and $\epsilon_{s}=12.56$ is the static dielectric constant.
Even though  the electron, hole and overlap terms in $\Phi_{ac}^{pz}(t)$ have an infrared divergence we 
can still estimate it in the same way the integrals for 
the deformation potential were estimated, i.e. via a saddle point approximation. The diverging part 
is eliminated since it does not depend on $\sigma$ and $\sigma_{z}$ and thus cancels out
when summing the electron hole and overlap term contributions. The result of the calculation is the following
(for details see Appendix \ref{subsec:appie})
\begin{eqnarray}
\label{eq:de}
\Phi_{ac}^{pz}(t)&=&  { C_{p} \over \hbar} \left \{ \ln{\bar{\sigma}_{z}^2 \over \sigma^{e}_{z} \sigma^{h}_{z}}
+ \sum_{j={e,h}} \left [{1 \over s_{j}}
\ln \left ( {1 - s_{j} \over 1 + s_{j}} \right ) 
+ {\pi \tau_{j}^2 e^{- {\left (t \over \tau_{j} \right )}^2 }  \over t^2 s_{j}} 
Erf\left({t \ s_{j} \over \tau_{j} }\right )\right ]\right  \} \nonumber \\
& - & {2 C_{p} \over \hbar} \left [{1 \over \bar{s}}
\ln \left ( {1 - \bar{s} \over 1 + \bar{s}} \right ) +
 {\pi {\bar{\tau}}^2 e^{- {\left (t \over \bar{\tau} \right )}^2} \over  t^2 \bar{s}} 
Erf\left({t \bar{s} \over \bar{\tau}} \right ) \right ] \ .
\end{eqnarray}
Again one must keep in mind that this result as the saddle point result for the deformation 
potential holds only for long times, i.e. $t \gg {p \over u}$. Typical dephasing time is 
given as in the case of the deformation potential coupling by 
$\tau_{dph}^j \simeq {2 \sigma^{j} \over u}$ ,i.e., time scales of the order of pico-seconds.
The remaining coherence for the case of no external electric filed is given by:
$g_{pz}= \Phi^{pz} (t \longrightarrow \infty)] =0.01$ and for the case of a strong field
is $g_{pz}=0.058$. It is instructive to compare these results with those obtained in Ref. \cite{Khun02},
to realize how strongly the piezoelectric coupling depends on the relative shape of the electron
and hole wave functions.

\section {Summary}
\label {sec:Summary}

We considered pure dephasing effects on a exciton due to scattering with optical and acoustical phonons.
Pure dephasing effects due to phonons are not effective in dephasing the exciton, i.e the non-diagonal term in the 
density matrix  (polarization) does not decay to zero but to some constant value
 which manifests itself as a sharp peak in the ZPL \cite{Khun02}.
 
Though the absorption spectra exhibits strongly non-Lorentzian line shapes \cite{Khun02}
still we were able to obtain intuitive expressions for
the typical dephasing times. The typical dephasing time due to coupling to acoustic phonons is given by $\tau_{dph}^{a} \approx
{2 \sigma_{e,h}\over u}$. This is simply the ratio of typical length scale for the electron/hole
wave function divided by the phonon sound velocity, which can be viewed as the time it takes an acoustic 
phonon to travel through the exciton wave function or as one over the energy of a phonon corresponding to the 
typical wave function length scale. The coupling to optical phonons strongly depends on the shape of the
wave function and goes to zero when the electron and hole have the same wave function. For a strong external electric
field, which spatially separates the electron and hole wave function we obtained the the typical time for
dephasing is well approximated by $\tau_{dph} \simeq {{\sigma}^2-{\sigma_{z}}^2 \over f}$, which can be roughly viewed
as the ratio of the QD anisotropy to the deviation from flatness of the optical phonon band.

We where also able to obtain simple expressions for the remaining coherence. For acoustic phonons coupled to the 
exciton through the deformation potential coupling it is given by ${C_{d} \over (2 \pi \sigma^{j})^2 \hbar}$ 
which also has an intuitive meaning. This expression is roughly the ratio of the  of the polaronic shift
(a measure of the electron phonon interaction strength),
assuming the main contribution to be from phonons of wave length $1/\sigma$, to the energy of such a typical phonon. 

It is interesting to note that the typical dephasing times do not depend on the strength of the
electron (hole) phonon interaction whereas the strength of the interaction strongly
determines the remnant coherence. We have also examined the effect of an external electric field on the dephasing process. 
Such a field was shown not to effect the typical decoherence time rather it effects the spectral weight of the zero 
phonon line, i.e. the remaining coherence (see Table. \ref{tab1}). The electric field controls the distance between the 
electron and hole wave functions and in this sense controls the strength of the phonon exciton interaction. The larger the 
field the less the cancellation between the electron and hole ``polaron clouds'' the more phonons are effective in decohering the exciton.
We have also shown the importance of using realistic wave functions for the confined electron and hole, the piezoelectric coupling
mechanism for acoustic phonons strongly depends on this difference as can be seen by comparing our results to those obtained
by Krummheuer and co-authors \cite{Khun02} which considered an infinite well confining potential
in the growth direction.   

Using typical scales for strong confinement QDs which are proposed as the basic building blocks
for quantum information processing in semiconductors (Ref. \cite {Biolatti00})
we have evaluated typical dephasing times and remnant coherence (see Table. \ref{tab1}). 
The typical dephasing time due to acoustic phonons was shown to
be of the order of pico-seconds whereas for optical phonons it is much larger on the order of $50-100 ps$.
When considering possible QIC implementations in QD both these 
dephasing times (for acoustic and optical phonons) should be compared with the typical time of exciting 
the exciton i.e. the Rabi frequency. Though more detailed calculations need to be conducted in which
the dephasing for the laser dressed \cite{Calarco}
states is considered. An important point which should be stressed is that the remnant coherence
which we calculated does not correspond directly to the fidelity of a possible QIP. One can use
the typical dephasing times in order to engineer adiabatic gating operations. Our calculation results 
suggest that properly engineering of the laser pulses can allow one to perform adiabatic 
(with respect to acoustical phonons) gating on the time scale of pico-seconds in which optical phonons will be gapped out. 

The author is grateful to T. Kuhn, F. Rossi and P. Zanardi for fruitful discussions
and acknowledges support through a Kreitman fellowship. 
\newpage

\appendix

\section{Dispersion relations}
\label{ap:dispersion}

As we mentioned before the dispersion relations has to be introduced (for optical phonons)
in order to get decoherence this is done in the following way. For optical phonons the dispersion relation is given as
\begin{equation}
\omega_{q}^2 = c_{1}^{op} + {\left [ (c_{1}^{op})^2 - c_{2}^{op} \sin^2 \left ({\pi q \over 2 k_{max}} \right )
\right ]}^{1 \over 2}
\end{equation}
where $ c_{1}^{op}= 1500 ps^{-2}$, $c_{2}^{op} = 2 \times 10^6 ps^{-4}$ for an fcc lattice $ k_{max}= {2 \pi \over a}$
($a$ is the lattice constant) and for GaAs ${\pi \over 2 k_{max}}= 0.143 nm$ 
(the values were obtained using  \cite{adachi94}). The optical frequency for $q=0$
is thus $\omega_{0}^{op}= 55 ps^{-1}$.

For the acoustic phonons the dispersion relation is given by:
\begin{equation}
\omega_{q}^2 = c_{1}^{ac} - {\left [ (c_{1}^{ac})^2 - c_{2}^{ac} \sin^2 \left ({\pi q \over 2 k_{max}} \right )
\right ]}^{1 \over 2}
\end{equation}

\section{Calculation of $\Phi(t)$}
\label{ap:calc}
\subsection{Optical phonons}
\label{subsec:optical}

In calculating Eq. (\ref{eq:disexp}) there are three different parts in $K(q)$,
it is convenient to use spherical coordinates, which leads to two 
integrals (one for the electron and one for the hole) of the following type 
\begin{eqnarray}
& & {B  \over 4 \pi^2 \hbar} \int_{0}^{\pi} d \theta \sin \theta \int_{0}^{\infty} d q 
\exp{\left [-q^2 (\sin^2 \theta \ \sigma^2 +\cos^2 \theta \ \sigma_{z}^2)\right ]} \nonumber \\
& = & {B \over 8 \pi^2 \hbar} \int_{0}^{\pi} d \theta \sin \theta 
\sqrt{\pi \over \sin^2 \theta \ \sigma^2 +\cos^2 \theta \ \sigma_{z}^2} \ .
\label{eq:spherical}
\end{eqnarray}
Integrating over angles one obtains
\begin{equation}
{B \sqrt{\pi} \over 8 \pi^2 \hbar} \int_{-1}^{1}{dx \over \sqrt{\sigma^2(1-x^2) + x^2\sigma_{z}^2}}=
{B \over 8 \pi^2 \hbar} \sqrt{\pi \over \sigma_{z}^2-\sigma^2} 
\ln \left [{\sigma_{z}+\sqrt {\sigma_{z}^2-\sigma^2} \over \sigma_{z} - \sqrt {\sigma_{z}^2-\sigma^2}} \right] \ .
\label{eq:angle}
\end{equation}
In the case of no electric field, i.e. $x_{0}=0$ the integral for the third part, the over-lap term, 
is also of the same sort.
When there is a strong electric field the integral for the over-lap term is a bit more involved 
\begin{eqnarray}
& &{B \over 4 \pi^2 \hbar}\int_{0}^{\pi} d \theta \sin \theta \int_{0}^{\infty} d q
J_{0}(q x_{0})\exp{\left [-q^2 (\sin^2 \theta \ \bar{\sigma}^2 +\cos^2 \theta \ \bar{\sigma}_{z}^2)\right ]}
\nonumber \\
& = & {B  \over 4 \pi^2 \hbar}\sqrt{\pi \over 2}\int_{0}^{\pi} d \theta \sin \theta \ 
{J_{0} \left[
{\sin^2 \theta \ x_{0}^2 \over 4 \ (\sin^2 \theta \ \bar{\sigma}^2 +\cos^2 \theta \ \bar{\sigma}_{z}^2)}
\right ] \over \sqrt{(\sin^2 \theta \ \bar{\sigma}^2 +\cos^2 \theta \ \bar{\sigma}_{z}^2)}} 
\exp\left [
- {\sin^2 \theta \ x_{0}^2 \over 4 \ (\sin^2\theta \ \bar{\sigma}^2 +\cos^2\theta \ \bar{\sigma}_{z}^2)}
\right ] \ .
\label{eq:apstrgfi}
\end{eqnarray}
This integral can be done using a saddle point approximation using the large parameter 
${x_{0} \over \sigma}$, the large contribution to the integral coming from the boundaries
$x=\pm 1$ giving
\begin{eqnarray}
&  &{B \over 4 \pi^2 \hbar}\sqrt{\pi \over 2} \int_{-1}^{1}
{dx \over \sqrt{x^2( \bar{\sigma}_{z}^2 - \bar{\sigma}^2) + \bar{\sigma}^2}}
J_{0} \left [ { (1-x^2)x_{0}^2 \over 4 [(1-x^2) \bar{\sigma}^2 + x^2 \bar{\sigma}_{z}^2]} \right ]
\times  \nonumber \\
& & \exp\left [{(1-x^2)x_{0}^2 \over 4 [(1-x^2) \bar{\sigma}^2 + x^2 \bar{\sigma}_{z}^2]} \right ] =
{B \over 2 \pi^2 \hbar}\sqrt{\pi \over 2}{\bar{\sigma}_{z} \over x_{0}^2} \ .
\label{eq:saddlept}
\end{eqnarray}

\subsection{Deformation potential}
\label{subsec:apdef}
Calculating $\Phi_{ac}^{def}(t)$ for acoustic phonons coupled to the exciton via deformation potential coupling
(Eq.\ref{eq:dephasing}) one can calculate the time independent part, for the electron and hole parts of $K(q)$,
analytically. First performing the integral for the integration for the $q$ plane,
\begin{eqnarray}
\label{eq:timeinde1}
& &{C_{d} \over {(2 \pi)}^3 \hbar}\int_{-\infty}^{\infty} d q_{z} \int_{0}^{\infty}  dq
{2 \pi q \over \sqrt{q^2 +q_{z}^2}}\ \exp{\left (-q^2 \sigma^2  -q_{z}^2 \sigma_{z}^2)\right ]} = \nonumber \\
& = &{C_{d} \over {2 \sigma^2 (2 \pi)}^2 \hbar} \int_{-\infty}^{\infty} {d q_{z} \over \sqrt{1+q_{z}^2}}
{1 \over \left [ 1+ q_{z}^2\ p^2 \right ]} \ ,
\end{eqnarray}
where the notation $p= {\left(\sigma_{z} \over \sigma \right)}$ has been used.
Integrating over $q_{z}$ one obtains
\begin{eqnarray}
\label{eq:timeinde2}
& &{C_{d} \over {2 \sigma^2 (2 \pi)}^2 \hbar} \int_{-\infty}^{\infty} {d q_{z} \over \sqrt{1+q_{z}^2}}
{1 \over \left [ 1+ q_{z}^2\ p^2 \right ]}
 \nonumber \\
& = & {C_{d} \over {\sigma^2 (2 \pi)}^2 \hbar}
{\tanh^{-1}{\left[ \sqrt{1- p^2}\right]} \over 
\sqrt{1 - p^2}} \ .
\end{eqnarray}
For the integration of time dependent part of the electron and hole parts of $K(q)$, it is 
convenient to use to use spherical coordinates. First performing the integration over the angles
\begin{eqnarray}
\label{eq:timede1}
& & {C_{d} \over {(2 \pi)}^2 \hbar}\int_{0}^{\pi} d \theta \sin \theta \int_{0}^{\infty}  
q \exp{\left [-q^2 ( \ \sigma^2 \sin^2 \theta  + \sigma_{z}^2 \ \cos^2 \theta )\right ]}
 \ e^{- \imath \omega_{q} t} = \nonumber \\
& = & {C_{d} \sqrt{\pi} \over {(2 \pi)}^2 \hbar \sigma_{z}} \int_{0}^{\infty} dq \
e^{-\imath q \left ( {u t \over \sigma_{z}} \right )}
e^{-{q^2 \over p^2}}
{Erf{ \left ({q\over p}  \sqrt{p^2 - 1}  \right )} \over \sqrt{\sigma_{z}^2 - \sigma^2}} \ .
\end{eqnarray}
Having scaled $q$ by $\sigma_{z}$
the integration over $q$ is now performed via the saddle point method, utilizing the typical anisotropy of 
self-assembled QD's, i.e. $p^{-2} \gg 1$. The saddle point given
by $q_{0}=-{\imath u t \sigma_{z} \over 2 \sigma^2}$, one obtains
\begin{eqnarray}
\label{eq:timede2}
& & {C_{d} \sqrt{\pi} \over {(2 \pi)}^2 \hbar \sigma_{z}} \int_{0}^{\infty} dq \
e^{-\imath q \left ( {u t \over \sigma_{z}} \right )}
e^{-{q^2 \over p^2}}
{Erf{ \left ( {q\over p}  \sqrt{p^2 - 1}  \right )} \over \sqrt{\sigma_{z}^2 - \sigma^2}} = \nonumber \\
& = & {C_{d} \over 4 \sigma^2 \pi \hbar} 
{e^{- u^2 t^2/4 \sigma^2} \over \sqrt{1- p^2}}
Erf{ \left ( {ut \over 2 \sigma} \left[ \sqrt{1- p^2} \right] \right )} 
\ .
\end{eqnarray}
This results holds only when $t$ is large enough, i.e. $t \gg {p \over u}$. 

The typical acoustic phonon with which the exciton interacts has a wave number of ${1 \over \sigma}$, using
this one can obtain the condition for the zero temperature approximation: $T< {u \hbar \over \sigma}$, i.e.
the temperature should be less than $2 K$. The condition can also be stated as the thermal phonon should
have a wave length much longer than the size of the QD or electron and hole wave functions 

\subsection{Piezoelectric coupling}
\label{subsec:appie}

The expression for the electron hole and overlap term contributions in $\Phi_{ac}^{pz}(t)$ have an infrared divergence. 
We will show, for the time independent part of $\Phi_{ac}^{pz}$, that the diverging part cancels out when summing these terms up.
The same procedure can be extended to the time dependent part. The integral for  $\Phi_{ac}^{pz}$ has three contributions
an electron a hole and a overlap term each of the following form 
\begin{eqnarray}
\label{eq:reg}
& &{C_{p} \over {(2 \pi)}^2 \hbar}\int_{0}^{\pi} \int_{0}^{\infty}  
{d q \over q}\left \{ \exp{\left [-q^2 (\sin^2 \theta \ \sigma^2 +\cos^2 \theta \ \sigma_{z}^2)\right ]}-
{1\over 1 + \Lambda^2 q^2} \right \} = \nonumber \\
& =& {C_{p} \over {(2 \pi)}^2 \hbar} \left [- \gamma + 2 +\ln {\Lambda \over \sigma_{z} } -{1 \over \sqrt{1- p^2}}
\ln \left ( {1 - \sqrt{1- p^2} \over 1 + \sqrt{1- p^2}} \right ) \right ]
\end{eqnarray}
where $\gamma \simeq 0.577216$ is Euler's constant. We have regularized the integral by subtracting from it, 
${1\over 1 + \Lambda^2 q^2}$, where  at the end of the calculation the limit $\Lambda \rightarrow \infty$ 
should be taken, but since there are three terms (corresponding to contributions from the electron hole and overlap)
summing them up the contributions which do not depend on $\sigma$
and $\sigma_{z}$ cancel out.

\newpage

\begin{table}
\caption{ Calculated typical dephasing times and remnant coherence due to coupling
with optical phonons (Fr\"{o}lich) and acoustic phonons (deformation potential and
piezoelectric). The remnant coherence is just the ratio of the final value of the absolute
value of the polarization squared to the initial value: $|P(t \rightarrow \infty )|^2 / |P(0)|^2$.}

\begin{tabular}{lccc}
Coupling mechanism & Dephasing time (ps) & 
\multicolumn{2}{c} {Remnant coherence} \\ 
  &  & (no external field) &(strong external field)  \\
Optical phonons (Fr\"{o}lich)) & $\sim 100$  & $0.992$ & $0.967$ \\
Acoustic (deformation potential) & $ \sim 1$ & $0.978 $ &$0.956$  \\
Acoustic (piezoelectric) &  $ \sim 1$  & $0.99$ & 0.944  \\
\label{tab1}
\end{tabular}
\end{table}


\begin{thebibliography}{aaaaaaaaaaaaaaaaaaaaa}
\bibitem{re:gan}
In the case of the GaN QDs the electric field due to strain is in the growth $z$-direction.
\bibitem{re:opt}
A strong electric field is used in the QIC scheme described in Ref. \cite {Biolatti00} in order to enhance
the dipole filed of the exciton.
\end{thebibliography}
\end{document}